\documentstyle[aps,prl,amsfonts,epsfig,multicol]{revtex}

\newcommand{\onecolm}{
  \end{multicols}
  \noindent\rule{0.5\textwidth}{0.1ex}\rule{0.1ex}{2ex}\hfill
}
\newcommand{\twocolm}{
  \hfill\raisebox{-1.9ex}{\rule{0.1ex}{2ex}}\rule{0.5\textwidth}{0.1ex}
  \begin{multicols}{2}
}

\begin{document}
\draft
\preprint{MSC-97-06}

\title{Quantum Phase Transitions in Capacitively Coupled
Two-Dimensional Josephson-Junction Arrays}
\author{Mahn-Soo Choi$^*$}
\address{Department of Physics and Pohang Superconductivity Center,
Pohang University of Science and Technology, Pohang 790-784, Korea}
\date{September 4, 1998}
\maketitle

\thispagestyle{empty}

\begin{abstract}
Quantum-phase transitions in two layers of ultrasmall Josephson
junctions, coupled capacitively with each other, are investigated.
As the interlayer capacitance is increased, the system at zero
temperature is found to exhibit an insulator-to-superconductor
transition.
It is shown that, unlike one-dimensional arrays with a similar
coupling configuration, the transition cannot be accounted for
exclusively by particle-hole pairs.
\end{abstract}

\pacs{PACS Numbers: 74.50.+r, 67.40.Db, 73.23.Hk}

\def\indicator{\noindent{\bf\P}\quad}
\newcommand\varD{{\cal D}}
\newcommand\varG{{\cal G}}
\newcommand\varS{{\cal S}}
\newcommand\bfq{{\bf q}}
\newcommand\bfr{{\bf r}}
\newcommand\bfv{{\bf v}}
\newcommand\bfm{{\bf m}}
\newcommand\avgl{\left\langle}
\newcommand\avgr{\right\rangle}
\newcommand\hatU{\widehat{U}}
\newcommand\tilU{\widetilde{U}}
\newcommand\tilC{\widetilde{C}}
\newcommand\CC{{\Bbb{C}}}

\begin{multicols}{2}

Capacitively coupled systems of charges have attracted significant
attention in recent years, raising the possibility of current drag
effects:
The current fed through either of the systems, owing to Coulomb
interaction, induces a secondary current in the other system.
Such a drag effect depends strongly on the dimensionality and the
structure of the system in its mechanism and behavior.
The current drag in two capacitively coupled two-dimensional (2D)
electron gases~\cite{Solomo89} was attributed to momentum-transfer
mechanism due to Coulomb scattering~\cite{Pogreb77} and fairly small
in magnitude.
By contrast, recent theoretical prediction~\cite{Averin91a} and
experimental demonstrations~\cite{Matter97,Delsin96} with two
capacitively coupled one-dimensional (1D) arrays of submicron metallic
tunnel junctions have shown that the primary and the secondary currents
are comparable in magnitude but opposite in direction in a certain
region of applied voltage.
In such tunnel junction systems, the current drag is attributed to the
transport of electron-hole pairs, which are bound by the electrostatic
energy of the coupling capacitance.
Lately, it has been suggested that the momentum-transfer mechanism can
also lead to the absolute current drag in 1D electron
channels coupled electrostatically with each other~\cite{Nazaro97}.
The current drag effects in capacitively coupled 2D arrays of tunnel
junctions has not been studied and will be examined in this work.

More interestingly, when the tunneling junctions are composed of ultrasmall
superconducting grains, the counter part of the electron-hole pair
becomes the pair of excess and deficit Cooper pair, which will be
simply called the particle-hole pair. 
Furthermore, in such ultrasmall Josephson junction systems the
competition between the charging energy and the Josephson coupling
energy is well known to bring the novel effects of quantum
fluctuations~\cite{Bradle84,Faziox91,BJKim95,mschoi98a}.
Combined with these quantum fluctuation effects, the pair transport
phenomena in coupled 1D Josephson junction arrays (JJAs) has recently
been proposed to drive an insulator-to-superconductor
transition~\cite{mschoi97p5}.

In this paper, two 2D arrays of ultrasmall Josephson
junctions, coupled capacitively with each other, are considered.
Quantum-phase transitions are examined at zero
temperature, focusing on the roles of the particle-hole pairs.
The system is transformed into two three-dimensional (3D) systems of
classical vortex loops, which are {\em topologically} coupled but
otherwise independent of each other.
The resulting model reveals that as the coupling capacitance
increases, in appropriate regions of parameters, the system exhibits
an insulator-to-superconductor transition.  Contrary to the 1D
counterpart with a similar coupling scheme~\cite{mschoi97p5},
the transition cannot be ascribed exclusively to the condensation of
the particle-hole pairs.
Accordingly, it is also remarked briefly that the accompanying drag of
supercurrents in the superconducting phase is not absolute in general.
In the vicinity of the transition, however, the particle-hole pairs
still play major roles, and therefore current drag can be large.

As a matter of fact, capacitively coupled 2D JJAs have been studied by
several authors, but in different context and different regions of
parameter space~\cite{Blante97}. 
Besides, the capacitive coupling should be distinguished from the
Josephson coupling such as considered in multi-layered
systems~\cite{Blatte94}.
The capacitively coupled JJAs can presumably be realized in experiment
by current techniques, which have already made it possible to fabricate
submicron metallic junction arrays with large inter-array
capacitances~\cite{Matter97,Delsin96} as well as large arrays of
ultrasmall Josephson junctions~\cite{Schonx90}.

\begin{figure}
\centerline{\epsfig{file=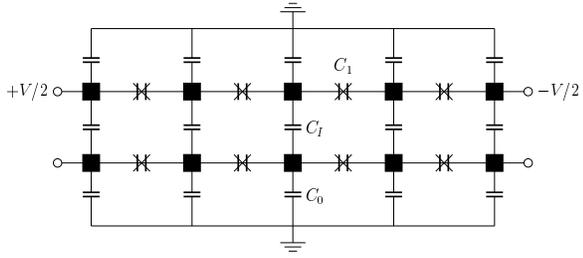,clip=,width=0.9\columnwidth}}
\caption{Schematic side view of the system.  Each chain in the figure
represents a 2D array.}
\label{fig:CCJJA}
\end{figure}

Each of the two arrays ($\ell=1,2$) of Josephson junctions considered
here is characterized by the Josephson coupling energy $E_J$ and the
charging energies $E_0{\equiv}e^2/2C_0$ and $E_1{\equiv}e^2/2C_1$,
associated with the self-capacitance $C_0$ and the junction
capacitance $C_1$, respectively (see Fig.~\ref{fig:CCJJA}).
The two arrays are coupled with each other by the capacitance $C_I$,
with which the electrostatic energy $E_I{\equiv}e^2/2C_I$ is
associated, while there is no Cooper-pair tunneling between the
arrays.
The intra-array capacitances are assumed to be so small
($E_0,E_1\gg{}E_J$) that, without the coupling, both arrays would each
be separately in the insulating phase~\cite{Faziox91}.
It is also assumed that the coupling capacitance is sufficiently large
compared with the intra-array capacitances; $C_I\gg{}C_0,C_1$.
In that case, the electrostatic energy of the particle-hole pair
($\sim{}E_I$) is much smaller than that of an unpaired charge
($\sim{}E_0,E_1$); the particle-hole pair, bound by the binding energy of
order of $E_0-E_I$ or $E_1-E_I$, is thus much favorable than the unpaired
charges.
For the most part, this work is devoted to the case of identical
arrays, but non-identical arrays will also be briefly discussed.

The system can be well described by the Hamiltonian
\begin{eqnarray}
H
& = & \frac{1}{4K}\sum_{\ell,\ell'}\sum_{\bfr,\bfr'}
  n(\ell;\bfr)\CC^{-1}(\ell,\ell';\bfr,\bfr')
  n(\ell';\bfr')
  \nonumber \\
& & \mbox{}
  - 2K\sum_\ell\sum_{<\bfr\bfr'>}
    \cos\left[\phi(\ell;\bfr)-\phi(\ell;\bfr')\right]
  ,
  \label{CCJJA:H}
\end{eqnarray}
where $\bfr\equiv(x,y)$ denotes the 2D
lattice vector in units of lattice constant,
the coupling constant has been defined by
$2K{\equiv}\sqrt{E_J/4E_I}$,
and the energy has been
rescaled in units of the Josephson plasma frequency
$\hbar\omega_p\equiv\sqrt{4E_IE_J}$.
The number $n(\ell;\bfr)$ of excess Cooper pairs and the phase
$\phi(\ell;\bfr)$ of
the superconducting order parameter on the grain at $\bfr$ in
array $\ell$ are quantum mechanically conjugate variables:
$
[n(\ell;\bfr),\phi(\ell';\bfr')]
{=}i\delta_{\bfr\bfr'}\delta_{\ell\ell'}
$.
The capacitance matrix in Eq.~(\ref{CCJJA:H}) takes the form
\begin{equation}
\CC
= \left( \begin{array}{cc}
    C & 0 \\ 0 & C
  \end{array} \right)
  + \frac{1}{2}\left( \begin{array}{rr}
    1 & -1 \\ -1 & 1
  \end{array} \right),
\end{equation}
where the submatrices $C(\bfr,\bfr')$  are defined by the Fourier
transform $\widetilde{C}(\bfq)=C_0{+}C_1\Delta(\bfq)$
with $\Delta(\bfq)\equiv\Delta(q_x)+\Delta(q_y)$;
$\Delta(k)\equiv{}2(1{-}\cos{k})$.  Here all the capacitances have
been rescaled by the relevant capacitance scale $2C_I$:
$C_0/2C_I{\to}C_0$ and $C_1/2C_I{\to}C_1$.

\begin{figure}
\centerline{\epsfig{file=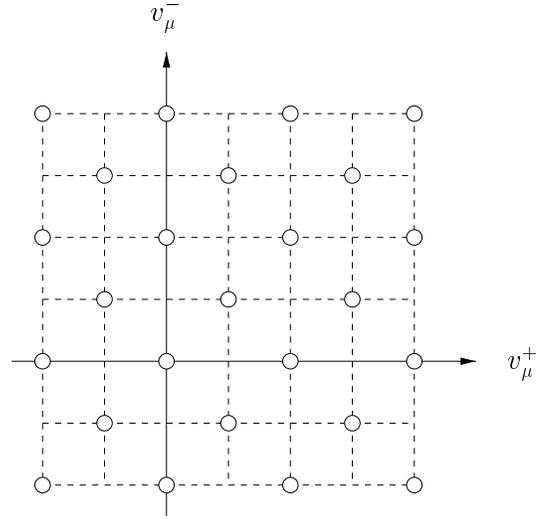,clip=,width=0.9\columnwidth}}
\caption{Topological coupling of $v^+_\mu$ and $v^-_\mu$.  At each
space-time position $\vec\bfr$, $(v^+_\mu,v^-_\mu)$ can take only half
of the elements in ${\bf{}Z{\times}Z}$ as depicted with open circles
in the figure.}
\label{fig:vpm}
\end{figure}

It is convenient to write the partition function of the system in
the imaginary-time path integral representation
\begin{equation}
Z
= \prod_{\ell,\bfr,\tau}
  \sum_{n(\ell;\bfr,\tau)}
  \int_0^{2\pi}d\phi(\ell;\bfr,\tau)\;
  \exp\left[ -\varS \right]
  \label{CCJJA:Z}
\end{equation}
with the Euclidean action
\begin{eqnarray}
\varS
& = & \frac{1}{4K}\sum_{\ell,\ell'}\sum_{\bfr,\bfr',\tau}
  n(\ell;\bfr,\tau)\CC^{-1}(\ell,\ell';\bfr,\bfr')
  n(\ell';\bfr',\tau)
  \nonumber \\
& & \mbox{}
  - 2K\sum_{\ell}\sum_{\bfr,\tau}\sum_j
    \cos\nabla_j\phi(\ell;\bfr,\tau)
  \nonumber \\
& & \mbox{}
  + i\sum_{\ell}\sum_{\bfr} n(\ell;\bfr,\tau)
    \nabla_\tau\phi(\ell;\bfr,\tau)
  ,
  \label{CCJJA:S}
\end{eqnarray}
where $\nabla_j$ ($j=x,y$) and $\nabla_\tau$ denote the difference
operators in the spatial and the imaginary-time
directions,  respectively, and the (imaginary-)time slice $\delta\tau$
has been chosen to be unity (in units of
$\omega_p^{-1}$)~\cite{end_note:2}.
The highly symmetric form of Eq.~(\ref{CCJJA:S}) with respect to space
and time makes it useful to introduce the space-time 3-vector notation
$\vec\bfr\equiv(\bfr,\tau)$, and analogous notations for all other
vector variables.
We then apply the Villain approximation~\cite{Josexx77} to rewrite the
cosine term as summation over integer fields
$\{m_x(\ell;\vec\bfr),m_y(\ell;\vec\bfr)\}
\equiv\{\bfm(\ell;\vec\bfr)\}$.
Further, with the aid of Poisson resummation formula~\cite{Josexx77}
and Gaussian integration, we rewrite the charging energy term as
summation over another integer field $\{m_\tau(\ell;\vec\bfr)\}$, to obtain
the partition function
\begin{equation}
Z
\sim \prod_{\ell;\vec\bfr}\sum_{\vec\bfm(\ell;\vec\bfr)}
  \int_{-\infty}^\infty d\phi(\ell;\vec\bfr)\;
  \exp\left\{ -\varS \right\}
  \label{tmp:1a}
\end{equation}
with
\onecolm  
\begin{eqnarray}
\varS
& = & K\sum_{\ell,\ell';\vec\bfr,\vec\bfr'}
    \CC(\ell,\ell';\bfr,\bfr')
    \delta_{\tau\tau'}\;
    \left[
      \nabla_\tau\phi(\ell;\vec\bfr)
      - 2\pi m_\tau(\ell;\vec\bfr)
    \right]
    \left[
      \nabla_\tau\phi(\ell';\vec\bfr')
      - 2\pi m_\tau(\ell';\vec\bfr')
    \right]
  \nonumber \\
& & \mbox{}
  + K\sum_{\ell;\vec\bfr}
    \left[
      \nabla\phi(\ell;\vec\bfr)-2\pi\bfm(\ell;\vec\bfr)
    \right]^2
  .
  \label{tmp:1b}
\end{eqnarray}
The variables
$\phi(\ell;\vec\bfr)$ and $\vec\bfm(\ell;\vec\bfr)$
can be usefully replaced by 
$\phi^\pm(\vec\bfr)\equiv\phi(1;\vec\bfr)\pm\phi(2;\vec\bfr)$ and
$\vec\bfm^\pm(\vec\bfr)\equiv\vec\bfm(1;\vec\bfr)\pm\vec\bfm(2;\vec\bfr)$,
respectively.
In this way, one decomposes the Euclidean action in Eq.~(\ref{tmp:1b})
into the sum $\varS=\varS^++\varS^-$ with $\varS^\pm$ defined by
\begin{eqnarray}
\varS^\pm
& = &
  + \frac{1}{2}K\sum_{\vec\bfr,\vec\bfr'}
    C^\pm(\bfr,\bfr')\delta(\tau,\tau')\;
    \left[
      \nabla_\tau\phi^\pm(\vec\bfr)
      - 2\pi m_\tau^\pm(\vec\bfr)
    \right]
    \left[
      \nabla_\tau\phi^\pm(\vec\bfr')
      - 2\pi m_\tau^\pm(\vec\bfr')
    \right]
  \nonumber \\
& & \mbox{}
  + \frac{1}{2}K\sum_{\vec\bfr}
    \left[
      \nabla\phi^\pm(\vec\bfr)
      - 2\pi\bfm^\pm(\vec\bfr)
    \right]^2
  .
  \label{S:pm}
\end{eqnarray}
\twocolm\noindent  
Here the new capacitance matrices $C^\pm(\bfr,\bfr')$ are defined via
the Fourier transforms $\tilC^+(\bfq)=\tilC(\bfq)$ and
$\tilC^-(\bfq)=1+\tilC(\bfq)$, respectively.
Now one follows the standard procedures~\cite{Korshu90,Faziox91} to
integrate out $\{\phi^\pm(\vec\bfr)\}$.
Then, apart from the irrelevant spin wave part, one can finally obtain
the 3D system of classical vortex lines, which is also decomposed into
two subsystems $H_V=H_V^++H_V^-$ with
\begin{equation}
H_V^\pm
= 2\pi^2 K\sum_{\vec\bfr,\vec\bfr'}\sum_\mu
    v_\mu^\pm(\vec\bfr)\;U_\mu^\pm(\vec\bfr-\vec\bfr')\;
    v_\mu^\pm(\vec\bfr')
    ,
    \label{H:3DVL:pm}
\end{equation}
where the interactions between
vortex line segments are defined via their Fourier transforms
\begin{eqnarray}
\tilU_\|^\pm(\vec\bfq)
& = & \frac{\tilC^\pm(\bfq)}
  {\Delta(\bfq)+\tilC^\pm(\bfq)\Delta(\omega)}
  \\
\tilU_\tau^\pm(\vec\bfq)
& = & \frac{1}
  {\Delta(\bfq)+\tilC^\pm(\bfq)\Delta(\omega)}
  .
\end{eqnarray}
Here the vortex lines $v^-_\mu$ are manifestation of the particle-hole
pairs, whereas $v^+_\mu$ stand for single particle
processes~\cite{mschoi97p5}.
Note that, in Eq.~(\ref{H:3DVL:pm}), the fields
$\{v^\pm_\mu(\vec\bfr)\}$ are subject to the constraint
$\vec\nabla\cdot\vec\bfv^\pm(\vec\bfr)=0$; i.e., all vortex lines
either form closed loops or go to infinity.
More importantly, it should also be noticed that the two fields
$v^+_\mu$ and $v^-_\mu$ cannot be independent of each other, since
$m_\mu(1;\vec\bfr)$ and $m_\mu(2;\vec\bfr)$ in Eq.~(\ref{S:pm}), and
hence $v_\mu(1;\vec\bfr)=[v^+_\mu(\vec\bfr)+v^-_\mu(\vec\bfr)]/2$
and $v_\mu(2;\vec\bfr)=[v^+_\mu(\vec\bfr)-v^-_\mu(\vec\bfr)]/2$, can
take only integer values.
As depicted with open circles in Fig.~\ref{fig:vpm},
$(v^+_\mu,v^-_\mu)$ at each $\vec\bfr$ can take only half of the
elements in the product set of integers ${\bf{}Z{\times}Z}$;
$v^+_\mu$ and $v^-_\mu$ are {\em topologically coupled} with each
other.
Contrary to the capacitively coupled 1D JJAs~\cite{mschoi97p5}, this
topological coupling plays crucial roles in the present case, which
will be discussed in more detail below. 

It is not difficult to understand the physics described by each of the
Hamiltonians $H_V^\pm$.  Unless $C_0=0$,
the length-scale dependence of the anisotropy factor
$\tilC^+(\bfq)=\tilC(\bfq)\ll1$
is screened out at length scales larger than $\sqrt{C_1/C_0}$,
and thereby $\tilU_\mu^+(\vec\bfq)$ is simply reduced to the highly
anisotropic current-like interaction:
$
\tilU_\|^+(\vec\bfq)
\simeq C_0/[\Delta(\bfq)+C_0\Delta(\omega)]
$
and
$
\tilU_\tau^+(\vec\bfq)
\simeq 1/[\Delta(\bfq)+C_0\Delta(\omega)]
$.
Such an anisotropic model has been studied in
Ref.~\onlinecite{Korshu90}, and is known to exhibit an anisotropic 3D 
transition which is associated with the disruption of
the vortex loops, at $K=K_c^+$ close to the 2D
Berezinskii-Kosterlitz-Thouless (BKT) transition point~\cite{Berezi71}:
$K_c^+\sim2/\pi$.  In the case of $C_0=0$, the vortex lines
$v_\mu^+$ even form 2D pancake vortices residing on decoupled 2D layers
with
$
\tilU_\|^+(\vec\bfq)
\simeq C_1/[1+C_1\Delta(\omega)]
$
and
$
\tilU_\tau^+(\vec\bfq)
\simeq 1/\Delta(\bfq)[1+C_1\Delta(\omega)]
$,
and the phase transition is precisely BKT-type.  In any case, the
system of vortex lines $v_\mu^+$ exhibits a phase transition at
$K=K_c^+\sim2/\pi$.  On the other hand, neglecting the 
anisotropy at short-length scales, $\tilU^-_\mu(\vec\bfr)$ are
isotropic in space-time:
$
\tilU_\|^-(\vec\bfq)
\simeq \tilU_\tau^-(\vec\bfq)
\simeq 1/[\Delta(\bfq)+\Delta(\omega)]
$.
In consequence, it follows that the system of vortex-lines $v_\mu^-$
exhibits the isotropic 3D XY-type phase transition at
$K=K_c^-\sim1/2\sqrt{2}$.
At this point, one might be tempted to conclude that, as $K$ is
increased, the total system $H_V$ might go through two successive
transitions, one at $K_c^-$ and the other at $K_c^+$, the first of
which would be ascribed to condensation of particle-hole
pairs~\cite{mschoi97p5}.
This scenario of successive transitions, however, should be tested
against the topological coupling discussed above between $v^+_\mu$ and
$v^-_\mu$.

For this goal, it is convenient to consider the subsystem
$\{v^*_\mu\}$ of $\{v^+_\mu\}$ satisfying $v^-_\mu=0$.
In this subsystem, $v^*_\mu(\vec\bfr)$ can take only even numbers, and
hence the phase transition could take place at
$K=K_*=K_c^+/4\sim1/2\pi$, which is substantially lower than $K_c^-$.
It means that vortices $v^*_\mu$ could be tightly bound even before
the vortices $v^-_\mu$ get bound, contradicting the assumption
$v^-_\mu=0$.
Consequently, it follows that actual phase transition should take
place at $K_c$ between $K_c^-$ and $K_c^+$, and cannot be accounted
for exclusively by $v^-$, i.e., by particle-hole pairs.
This is distinctive different from the case of the capacitively
coupled 1D chains~\cite{mschoi97p5}, where the vortices $v^*$ in
analogous subsystem always form a plasma of free vortices regardless
of $K$ in the presumed configuration $C_0,C_1\ll1$ and the topological
coupling is thus irrelevant; the free vortices $v^*$ completely screen
out the interaction among vortices $v^+_\mu$.

Nevertheless, it is evident that any corrections of the vortex lines
$v^+_\mu$ in the vicinity of $K_c$ is exponentially small in the
creation energy $\mu_c^+$ of the smallest vortex loops (or
nearby pancake vortex-antivortex pairs when $C_0=0$):
$\mu_c^+\sim{}K_c\pi$~\cite{Berezi71}. 
In particular, the shift of the transition point $K_c$ with respect to
$K_c^-$ can be estimated by
\begin{equation}
(K_c - K_c^-)/K_c^-
\sim e^{-2\mu_c^+}
\sim 0.1
  ,
\end{equation}
where the factor $2$ in the exponent is due to the topological
coupling.

Now I examine briefly and qualitatively the current drag effects in the
superconducting phase, by means of the linear response
$\sigma^{}_{\ell\ell'}(\omega)$ of the current in the array $\ell$ to the
voltage applied across the array $\ell'$ (see Fig.~\ref{fig:CCJJA}):
\begin{equation}
\sigma^{}_{\ell\ell'}(\omega)
= \frac{1}{i\omega}\lim_{\bfq\to0}
  \widetilde\varG_{\ell\ell'}(\bfq,i\omega'\to\omega+i0^+),
  \label{sigma:def}
\end{equation}
where $\widetilde\varG_{\ell\ell'}$ is the Fourier transform of the
imaginary-time Green's function
$
\varG_{\ell\ell'}(\bfr,\tau)
= \avgl T_\tau [I(\ell;\bfr,\tau)I(\ell';{\bf0},0)] \avgr
$
with the time-ordered product $T_\tau$
and the current operators
$I(\ell;\bfr)\equiv\sin\nabla_x\phi(\ell;\bfr)$ (since the system is
isotropic in $x$- and $y$-direction, only the current in the
$x$-direction is considered here).
Due to the symmetry between the two arrays, it follows that 
$
\sigma_{11}(\omega)
= \left[\sigma_+(\omega)+\sigma_-(\omega)\right]/2
$
and
$
\sigma_{21}(\omega)
= \left[\sigma_+(\omega)-\sigma_-(\omega)\right]/2
$,
where $\sigma_\pm$ are defined in a manner analogous to
Eq.~(\ref{sigma:def}) with
$I^\pm(x)\equiv{}I(1;x)\pm{}I(2;x)$.
According to the discussion above on the phase transition,
at $K>K_c$, both $\sigma_+$ and $\sigma_-$ show superconducting
behavior:
$
\sigma_\pm(\omega)
= \sigma_\pm^0\:\delta(\omega)
$
$(\omega \ll 1)$,  
which means that the drag of supercurrents along the two arrays is not
perfect in general.  However, in the vicinity of the phase transition,
where $\sigma_+^0\ll\sigma_-^0$, the currents in two arrays can be
comparable in magnitude.
This suggests the following: The particle-hole pair is not so tight as in
the 1D case, distributing over a few lattice constants.  Yet it is still
energetically favorable enough to play significant (if not exclusively
crucial) roles in the phase transition and the transport.

Before concluding, I remark briefly on non-identical arrays.  The
difference in the intra-array capacitances leads to additional
coupling between the vortices $v^+_\mu$ and $v^-_\mu$ with the
coupling strength proportional to the difference.  The arguments on
identical arrays therefore remain valid qualitatively as long as
$
\left|\tilC(1;\bfq)-\tilC(2;\bfq)\right|
\ll\left|\tilC(1;\bfq)+\tilC(2;\bfq)\right|
$.
The difference in Josephson coupling energy, on the other hand, can be
effectively incorporated in the capacitance difference by
renormalizing the parameters, since all the effects considered in this
work depends only on the relative strength of the Josephson coupling
energy and the charging energies.

In conclusion, quantum phase transitions in two capacitively coupled
2D JJAs have been investigated.
In particular, it has been found that as the coupling capacitance
increases, in appropriate parameter ranges ($E_J/E_0,E_J/E_1\ll1$;
$E_J/E_0,E_J/E_1\ll{}E_J/E_I<\infty$), the system exhibits an
insulator-to-superconductor transition.
Contrary to the capacitively coupled 1D chains, the transition cannot
be account for exclusive by the condensation of particle-hole pairs.
Accordingly, the drag of supercurrents along the two arrays is not
absolute in general.

I am grateful to M.~Y. Choi, S.-I. Lee, and J.~V. Jos\'e for valuable
discussions.  This work was supported by the Ministry of Science and
Technology of Korea through the by the Creative Research Initiative
Program.


\end{multicols}

\newpage
\begin{multicols}{2}

\narrowtext 

\end{multicols}

\end{document}